\documentclass[%
 aip,
 amsmath,amssymb,
preprint,
]{revtex4-1}

\usepackage{graphicx}
\usepackage{dcolumn}
\usepackage{bm}

\usepackage[utf8]{inputenc}
\usepackage[T1]{fontenc}
\usepackage{mathptmx}
\usepackage{etoolbox}
\usepackage{xcolor}
\usepackage{soul}
\usepackage{hyperref}

\makeatletter
\def\@email#1#2{%
 \endgroup
 \patchcmd{\titleblock@produce}
  {\frontmatter@RRAPformat}
  {\frontmatter@RRAPformat{\produce@RRAP{*#1\href{mailto:#2}{#2}}}\frontmatter@RRAPformat}
  {}{}
}%
\makeatother
\begin{document}


\title[Exploring the Spin Dynamics of a Room-Temperature Diamond Maser using an Extended rate Equation Model]{Exploring the Spin Dynamics of a Room-Temperature Diamond Maser using an Extended Rate Equation Model}
\author{Yongqiang Wen}
\author{Philip L. Diggle}%
\author{Neil McN. Alford}%
\author{Daan M. Arroo}
\email{d.arroo14@imperial.ac.uk}
\affiliation{
Department of Materials, Imperial College London, Exhibition Road, London SW7 2AZ, United Kingdom}
\affiliation{London Centre for Nanotechnology, 17-19 Gordon Street, London WC1H 0AH, United Kingdom
}

\date{\today}

\begin{abstract}
Masers - the microwave analogue of lasers - are coherent microwave sources that can act as oscillators or quantum-limited amplifiers. Masers have historically required high vacuum and cryogenic temperatures to operate, but recently masers based on diamond have been demonstrated to operate at room temperature and pressure, opening a route to new applications as ultra-low noise microwave amplifiers. For these new applications to become feasible at a mass scale, it is important to optimise diamond masers by minimising their size and maximising their gain, as well as the maximum input power of signals that can be amplified. Here, we develop and numerically solve an extended rate equation model to present a detailed phenomenology of masing dynamics and determine the optimal properties required for the cavity, resonator and gain medium in order to develop portable maser devices. We conclude by suggesting how the material parameters of the diamond gain media and dielectric resonators used in diamond masers can be optimised and how rate equation models could be further developed to incorporate the effects of temperature and nitrogen concentration on spin lifetimes.
\end{abstract}

\maketitle

\section{Introduction}

Due to their long spin relaxation times, negatively-charged nitrogen-vacancy  defect centres (NV$^{-}$) in diamonds \cite{zaitsev,doherty2013nitrogen} have been widely used as room-temperature quantum materials for quantum metrology \cite{barry2020sensitivity,graham2023fibercoupled,patel2020subnanotesla}, communications and quantum information processing \cite{hensen2015loophole,ruf2021quantum,bradley2019ten,weber2010quantum}. 
Recently, their use has been extended to the microwave analogue of lasers, known as masers. Although masers  are unmatched for low-noise microwave amplification, their use has historically been limited to niche applications in deep space communications and radio astronomy due to the need for high vacuum and cryogenic temperatures required to allow them to operate. The demonstration of room-temperature masers based on organic \cite{oxborrow2012room,ng2021quasi,Ng2023,Attwood2023} and inorganic \cite{jin2015proposal,breeze2018continuous,sherman2021performance,blank2022,zollitsch2023maser} gain media thus opens up a path to the widespread use of masers as ultra-low noise microwave amplifiers, which in addition to existing applications in radio astronomy and communications technology are of increasing interest for the low-power microwave signal processing essential to many quantum technologies \cite{arroo2021perspective}. In this latter regard, masers may offer an advantage over Josephson Parametric Amplifiers currently used for such applications, which have a limited dynamic range and are easily saturated \cite{roy2016introduction}.

The NV$^-$ defect is an $S=1$ spin centre that can be described using a simple Hamiltonian consisting of a ground state triplet and an excited-stated triplet. NV$^-$ defects can be photoexcited to the excited state triplet using laser pulses with wavelengths at or below 637 nm, from which they may relax either through a spin-conserving radiative decay or through a spin-selective non-radiative decay whose net effect is to preferentially transfer electrons from the excited $m_s = \pm 1$ states to the $m_s = 0$ ground state via a metastable singlet state, as seen in Fig. \ref{fig:energyLevels}. These spin-selective transitions mean that NV$^-$ centres can be optically controlled and read out. The defect exhibits a long $T_1$ spin-lattice relaxation time (up to 6 ms \cite{jarmola2012temperature}) and $T_2$ spin-spin relaxation time (1.8 ms in isotopically purified $^{12}$C diamond\cite{bala}, 0.7 ms for diamonds that have a natural abundance of $^{13}$C isotopes \cite{stanwix2010coherence}) at room temperature. It is these properties that allow an ensemble of NV$^-$ defects to be used as a gain medium for a room-temperature maser, as well as for lasers \cite{savvin2021nv}. The maser threshold pump rate is proportional to $(T_1(C-1))^{-1}$, where $C=4g^2N/(\kappa_c\kappa_s)$ is the cooperativity, $g$ is the coupling strength between each spin and the cavity field, $N$ is the number of spins, $\kappa_c$ is cavity loss rate, $\kappa_s$ is the dephasing rate of the spins and $T_1$ is the spin-lattice relaxation time \cite{breeze2018continuous} The masing frequency is determined by the energy difference between the $m_s=0$ and $m_s=-1$ level, thus, the masing frequency can be tuned via an external magnetic field through the Zeeman effect \cite{jelezko2006single, barry2020sensitivity}.

Despite impressive experimental demonstrations and characterisation of room-temperature diamond masers \cite{breeze2018continuous, blank2022,zollitsch2023maser}, the widespread adoption of diamond masers will remain challenging until technical difficulties such as making diamond gain medium with sufficient amount of NV$^-$ centres, high Q-factor cavity, strong and homogeneous external magnetic field and alignment with NV$^-$ axis are addressed \cite{Patel2021}. Here we use extended rate equation simulations to investigate how maser performance depends on the properties of the diamond gain medium and the dielectric resonator.

Specifically, we extend the standard first-order rate equations sometimes used to model the photophysics of NV$^{-}$ centres \cite{sherman2021performance} to a system of coupled second-order equations that incorporate the number of photons in the cavity. The extended second-order rate equations incorporate the number of photons in the cavity and allow the time-evolution of the populations of NV$^{-}$ centres in different energy states and the photon  number to be simulated explicitly.
The rate equations are employed to model the spin dynamics of the NV$^-$ centres under optical pumping at a wavelength of 532 nm. The results show masing dynamics with various parameters such as the loaded Q-factor of the cavity and the mode volume of the resonator. As well as aiding the design of room-temperature diamond masers, the extended rate equations presented will be useful for describing microwave mode-cooling devices \cite{ng2021quasi} and quantum heat engines based on NV$^-$ centres in diamond \cite{klatzow2019experimental}.

\section{Theoretical model}
\begin{figure*}
    \centering
    \includegraphics{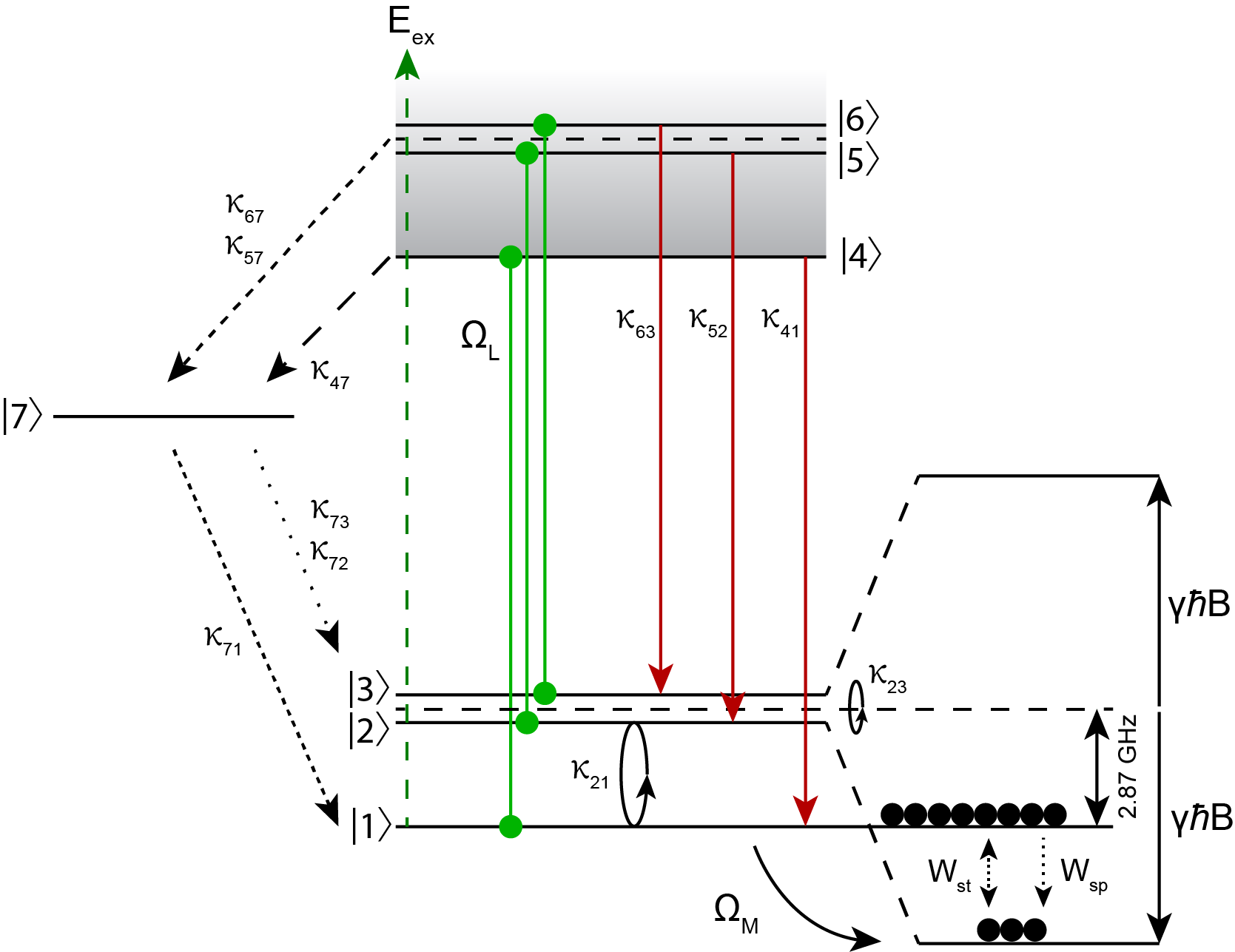}
    \caption{This diagram shows the energy levels, decay rate and pumping rate of the 7-level model. The excited phonon bands are shown in a green area in levels 4, 5 and 6. The green dashed line represents the 532 nm laser that pumps the population to the phonon bands, whereas the green solid line is the transition represented by the pump strength $\Omega_l$. The $\Omega_m$ indicates microwave excitation.
    $i$ indicate the levels, the $\kappa_{ij}$ indicates the  depopulation rate from level $i$ to level $j$, $\gamma \hbar B$ is the Zeeman splitting, $W_{st}$ and $W_{sp}$ is the stimulated emission rate and the spontaneous emission rate. The rates in the dashed line are larger than the rates in the dotted lines, and the more dashes or dots in the lines the higher the rate. }
    \label{fig:energyLevels}
\end{figure*}

The populations of the electronic energy levels of the NV$^{-}$ can be described by a 7-level model in Fig. \ref{fig:energyLevels}, where Levels 1 and 4 correspond to the $m_s=0$ states, and levels 2, 5 and 3, 6 correspond to the $m_s=-1$ and $m_s=+1$ respectively. The spin dynamics of different states under an optical excitation on NV$^-$ spin defects are described by rate equations based on a 7-level model based on an extended Bloch equation formalism \cite{jensen2013light,popa2006pulsed}. 
In this model, a laser pumps electrons from the ground-state sublevels 1, 2 and 3 respectively to the excited-state sublevels 4, 5 and 6 all at the same optical pumping strength $\Omega_l$.
The laser pump strength is defined below.
\begin{equation}
    \Omega_l=\mathcal{E}\mathbf{D_{ij}}/\hbar 
\end{equation}

where $\mathcal{E}$ is the electric field of the laser and $\mathbf{D_{ij}}$ is the dipole moment of the induced transition between level $i$ and level $j$, such as the ground triplet states and the excited triplets.
When a green laser at 532 nm is used to excite the NV$^-$ spin defects from the 1, 2 and 3 spin sub-levels of the ground state, the defects ``overshoot'' the transition to the corresponding excited-state sublevels before relaxing non-radiatively to the ground phonon state of the excited state indicated by levels 4, 5 and 6. Since the phonon relaxation is generally fast compared to the optical pump rate, the transitions $1 \to 4$, $2 \to 5$ and $3 \to 6$ are given a common rate defined by $\Omega_l$. From the excited states, the NV$^-$ centres can either decay radiatively through spontaneous emission of a photon at 637 nm (indicated by red lines with respective decay rates) or non-radiatively through the intersystem crossing, which we represent with a singlet state 7.

The populations of excited states then decay back to the corresponding ground-state sublevels with decay rates, $\kappa_{41}$, $\kappa_{52}$ and $\kappa_{63}$. Additionally, the population of the excited levels also decay to the singlet state, level 7, with rates, $\kappa_{47}$, $\kappa_{57}$ and $\kappa_{67}$. The populations of level 5 and level 6 are much more likely to cross to the singlet state, $\kappa_{57}$, $\kappa_{67} \gg \kappa_{47}$. The population of level 7 decays back to level 1 with a rate, $\kappa_{71}$, and to level 2 and level 3 with rates, $\kappa_{72}$ and $\kappa_{73}$. Furthermore, the populations of level 1, level 2 and level 3 are roughly equal at room temperature and in thermal equilibrium, due to the small energy differences between each level. Any spin polarisation will be depolarised at the longitudinal relaxation rate, $\kappa_{21}$. The rate equations thus form equations of motion for density matrix elements expressed with respect to the 7-level Jaynes-Cummings Hamiltonian.

\begin{multline}
    \mathcal{H} = \sum_{i=1}^{7} \hbar \omega_i|i \rangle \langle i|\\
 +\hbar \Omega_m  \left(|1\rangle \langle 2|+|2 \rangle\langle 1|\right)+\hbar \Omega_l  \sum_{\substack{i=1,2,3\\j=4,5,6}} \left(|i\rangle \langle j|+|j \rangle\langle i|\right)\\
\end{multline}
where $\omega_i$ is the frequency of the $i$th energy level, $|i\rangle$, of the NV$^-$ centres, $\omega_m$ is the microwave excitation strength, $\Omega_l$ is the coupling strength of the optical field, which is the pump strength that is determined by the electric field amplitude of the excitation laser at the NV$^-$ and the transition dipole moment of the ground state to the excited states of the NV$^-$ centres. 

The rate equations can be derived by taking the matrix elements of the Hamiltonian. The decay rates between the levels are added in diagonal terms, representing the populations, whereas the dephasing terms are added in the non-diagonal terms, representing the coherences. The photon number is treated as a second-order rate equation that depends on the number of thermal photons, and the spontaneous and stimulated transition rates. These additional terms allow the rate equations to be extendable, such as the operating frequency through the effect of the spontaneous and stimulated emission rate, as well as the dephasing and detuning terms in the coherence terms that describe the coherence between the spins and the cavity fields and the spectral properties of the model. 

The full matrix equations are shown in Appendix A. Using rates reported in \cite{klatzow2019experimental}, they were solved numerically using the RK45 method and an implicit Backward Differentiation method \cite{vetterling2002numerical} to simulate the time evolution of a diamond maser.

\begin{table*}
\caption{\label{tab:Parameters}Transition rates used to describe the time evolution of the ensemble of NV$^-$ centres in a cavity to produce the results presented in Figures 3-8. The values are based on experimental results from Ref. \cite{klatzow2019experimental}.}

\begin{ruledtabular}
\begin{tabular}{cccc} 
\hline
&Parameters &Values (MHz)  \\ [0.5ex] 
\hline
&$\kappa_{41}$=$\kappa_{52}$  =$\kappa_{63}$ & 65.9\\ 
&$\kappa_{57}$=$\kappa_{67}$ & 53.3  \\
&$\kappa_{47}$ & 7.9\\
&$\kappa_{71}$ & 0.98\\
&$\kappa_{72}$ =$\kappa_{73}$& 0.73\\
&$\kappa_{21}$ =$\kappa_{23}$& 2$\times 10^{-4}$ \\[1ex] 
\hline
\end{tabular}
\end{ruledtabular}
\end{table*}

In general, the dipole moment for the NV$^-$ ground and excited states can depend on the amount of strain in particular diamond samples, but for these simulations, we use a dipole moment of 5.2 C$\cdot$m, estimated from the fluorescence lifetime \cite{alkauskas2014first}. The electric field amplitude is determined by the laser pump power available at the NV$^-$ centres with respect to the absorption dipole direction, laser polarisation and laser spot size.

\section{Maser simulation}

The 7-level rate equations can be used to predict masing operation when a magnetic field above about $102.5$ mT, is applied such that the $m_s=-1$ state is below the $m_s=0$ state. The optical pumping polarises the $m_s=0$ state, thus, population inversion is achieved. The population will interchange between $m_s=0$ to $m_s=-1$ via spin-lattice relaxation to maintain thermal equilibrium, at a rate $\kappa_{12}$, and via spontaneous emission and stimulated emission which results in photon emission processes. The population will be excited to the $m_s=0$ state from $m_s=-1$ via spin-lattice relaxation and stimulated absorption. In the microwave (MW) spectrum, the spin-lattice relaxation rate, on the order of $10^2$ Hz, is much higher than the spontaneous emission rate in free space, which is on the order of $10^{-13}$ Hz \cite{purcell1995spontaneous}. The excited spins preferentially relax through the spin-relaxation mechanism by transferring energy to the lattice. The spontaneously emitted MW photons are negligible in free space.
The spontaneous emission rate is related to the Einstein coefficients $\mathbf{A}$ and $\mathbf{B}$, which phenomenologically describe the rate of the spontaneous and stimulated emission processes and can be boosted through the Purcell effect. The magnitude of this boost is proportional to the loaded Q factor and inversely proportional to the mode volume $Q/V_m$, which appears in the figure of merit of maser systems known as the cooperativity $C$, where $C>1$ can be taken as a necessary but not sufficient condition for masing\cite{breeze2018continuous}. 

The transition rate from one state $i$ to another state $j$ is determined\cite{berestetskii1982quantum} by the magnetic dipole matrix element of the corresponding two states, $\mathbf{d_{ij}}=\langle i|\mathbf{H}\cdot \mathbf{S}|j\rangle$, where $\mathbf{H}$ is the magnetic field and $\mathbf{S}$ is the spin vector, and the density of states of the electromagnetic field, $\rho(\omega)$ as $W \propto \dfrac{ |\mathbf{d_{ij}}|^2 \rho(\omega)}{\hbar }$. If the emitter is placed in a cavity, the density of state will alter, and the emission rate will either increase or decrease. 
The change in rate is determined by the Purcell factor, $F=\dfrac{3Qc^3}{2\pi V_m \omega^3 n^3}$, where $Q$ is the cavity loaded Q factor, $c$ is the speed of light, $\omega$ is the frequency, $V_m$ is the mode volume and n is the refractive index in the cavity\cite{purcell1995spontaneous}. The spontaneous and stimulated emission rate is directly proportional to the Purcell factor.
To enhance the spontaneous emission rate in a given frequency, a high Q factor and small mode volume are desired. The Purcell factor is sensitive to the frequency, $F \sim \omega^{-3}$, hence it might be beneficial to operate a maser at low frequency if the mode volume is not scaled up rapidly compared to the frequency. The stimulated emission and absorption rate are related to the spontaneous emission rate according to the Einstein coefficients.

 If the emitters are placed in a cavity, the density of states of the electromagnetic field is altered by reducing the number of states, thus increasing the spontaneous emission rate. 
The stimulated emission rate is $\kappa_{st}=\kappa_{sp}u(\nu)c^3\pi^2/h \omega^3$, where $u(\nu)$ is the energy density of the MW field and $\kappa_{sp}$ is the spontaneous emission rate.
In the masing simulation, the initial photon number in the cavity is determined by the thermal photon number at thermal equilibrium, $n_{th}=(e^{ \hbar\omega/k_B T}-1)^{-1}$. At 9.22 GHz and room temperature, the $n_{th}$ is approximately 661. MW photons are generated via spontaneous emission and trigger stimulated emission and absorption.  
For a realistic emission, dephasing and decoherence will contribute to the rate, which replaces the density of state by the linewidth function, $g(\omega)$. Given that the magnetic field of the MW is $\mathbf{H}$,the stimulated transition rate can be expressed \cite{vuylsteke1960elements}:
\begin{equation}
W=\dfrac{1}{4}\left(\dfrac{2\pi g \mu_0 \mu_B \mathbf{H}}{h}\right)^2g(\omega) \sigma^2
\end{equation}
where $\sigma^2=\dfrac{\mathbf{H}^*\cdot\sigma \sigma^* \cdot \mathbf{H}}{\mathbf{H}^*\cdot \mathbf{H}}$ can be interpreted as the normalised imaginary part of the magnetic susceptibility. For a two-level system interacting with a microwave field with positive circular polarisation, the maximum value of $\sigma^2$ is 0.5, while a microwave field with negative circular polarisation gives 0 as it will oppose the spin precession \cite{siegman1964microwave}.
Provided that the magnetic field energy density is $H^2=nhf\mu_0/V_m$, where $V_m$ is the mode volume and $n$ is the number of photons in the mode and $W=Bn$, the Einstein $\mathbf{B}$ coefficient at resonance is given by \cite{vuylsteke1960elements}:
\begin{equation}
\mathbf{B}=\dfrac{1}{4}\left(\dfrac{2\pi g \mu_B }{h}\right)^2\dfrac{\mu_0T_2^*hf\eta}{V_m}
\end{equation}
where the filling factor $\eta$ and the mode volume $V_m$ are respectively defined by
\begin{equation}
\eta=\dfrac{\int_{sample} |\mathbf{H}|^2 dV}{\int_{mode} |\mathbf{H}|^2 dV}
\end{equation}
\begin{equation}
V_m=\dfrac{\int_{sample} |\mathbf{H}|^2 dV}{|\mathbf{H}_{max}|^2}
\end{equation}
The higher the filling factor, the more strongly the sample interacts with the cavity mode.

For the results presented in the following section, the parameters from Table \ref{tab:Parameters} and Table \ref{tab:variables} are used. The finite-element software CST Microwave Studio is used to estimate the mode volume and the filling factor, using equations (4) and (5). In the CST simulation, we model a sapphire resonator with an inner diameter of 5 mm and an outer diameter of 10 mm on a sapphire post of 12 mm high in a copper cavity with an inner diameter of 31 mm and a height of 30 mm. The energy density of the magnetic field of the TE$_{01\delta}$ mode of the cavity is shown in Fig. 2. 
The $B$  coefficient is approximately 1.98$\times 10^{-6}$ s$^{-1}$. The spontaneous rate is given by \cite{siegman1964microwave}:
\begin{equation}
A=\frac{8\pi^3\mu_0f^3(g\mu_B)^2}{3hc^3}
\end{equation}
At 9.22 GHz, the spontaneous emission rate is about $1.57\times 10^{-12}$ s$^{-1}$, which is much slower than the stimulated transition rate and the spin relaxation rate. Knowing both the Einstein $\mathbf{A}$ and $\mathbf{B}$ coefficients, the stimulated emission rate from level $i$ is simply $W_{st}=B \rho_{ii} N(\omega)$, where $N(\omega)$ is the number of photons at frequency $\omega$, and the spontaneous emission rate from level $i$ is $W_{sp}=A \rho_{ii}$.

\begin{figure}
    \centering
    \includegraphics[width=1.0\columnwidth]{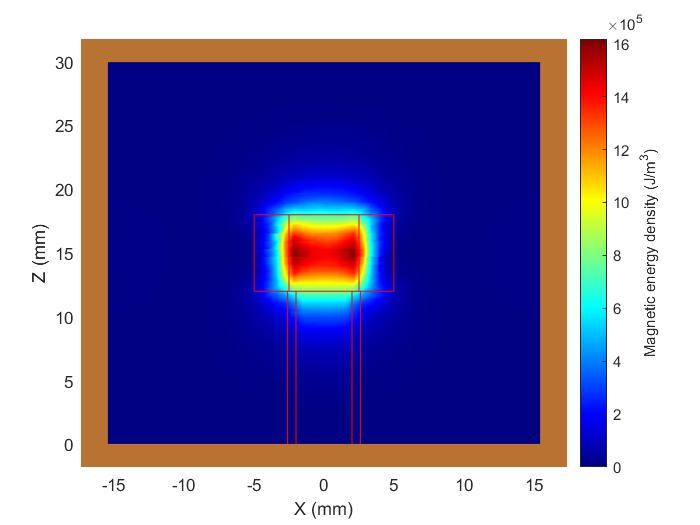}
    \caption{The energy density of the magnetic field of the TE$_{01\delta}$ mode of a cylindrical copper cavity simulated using CST Microwave Studio. The red outlines show the boundaries of a sapphire resonator and post. The walls of the copper cavity are shown as the enclosure that bounds the image.}
    \label{fig:energy-density}
\end{figure}

\begin{table}
\caption{\label{tab:variables} Variable parameters used in the rate equations to describe the time evolution of the ensemble of NV$^-$ centres in a cavity. The number of available NV$^-$, the maser frequency, $\omega$, the dephasing time, $T_2^*$ and the $Q$ factor is from \cite{breeze2018continuous}. The other parameters are obtained via CST Microwave studio simulations.}

\begin{ruledtabular}
\begin{tabular}{cccc} 
\hline
&Parameters &Values   \\ [0.5ex] 
\hline
&$N$ & 4$\times$ 10$^{13}$  \\ 
&$\omega$ & 2$\pi\times$ 9.22 GHz  \\ 
&$T_2^*$ & 0.5 $\mu$s  \\ 
&$V_m$ & 0.2 cm$^3$  \\ 
&$Q$ & 30,000  \\ 
&$\eta$ & 0.05  \\ 
&$B$ & 432 mT    \\
&$\Omega_l$ & 10 Hz \\ 
\hline
\end{tabular}
\end{ruledtabular}
\end{table}
The optical pumping interaction strength is 10 Hz.
The initial population is set by the Boltzmann distribution, $e^{\hbar \omega_m/k_BT}$, and the initial number of photons is $(e^{\hbar \omega_m/k_BT}-1)^{-1}$, where $k_B$ is the Boltzmann constant, $T$ is the temperature and the $\omega_m$ is the frequency of the cavity field, which is the frequency difference between the $m=0$ and $m=-1$ states.   

\begin{figure}
    \centering
    \includegraphics[width=.9\columnwidth]{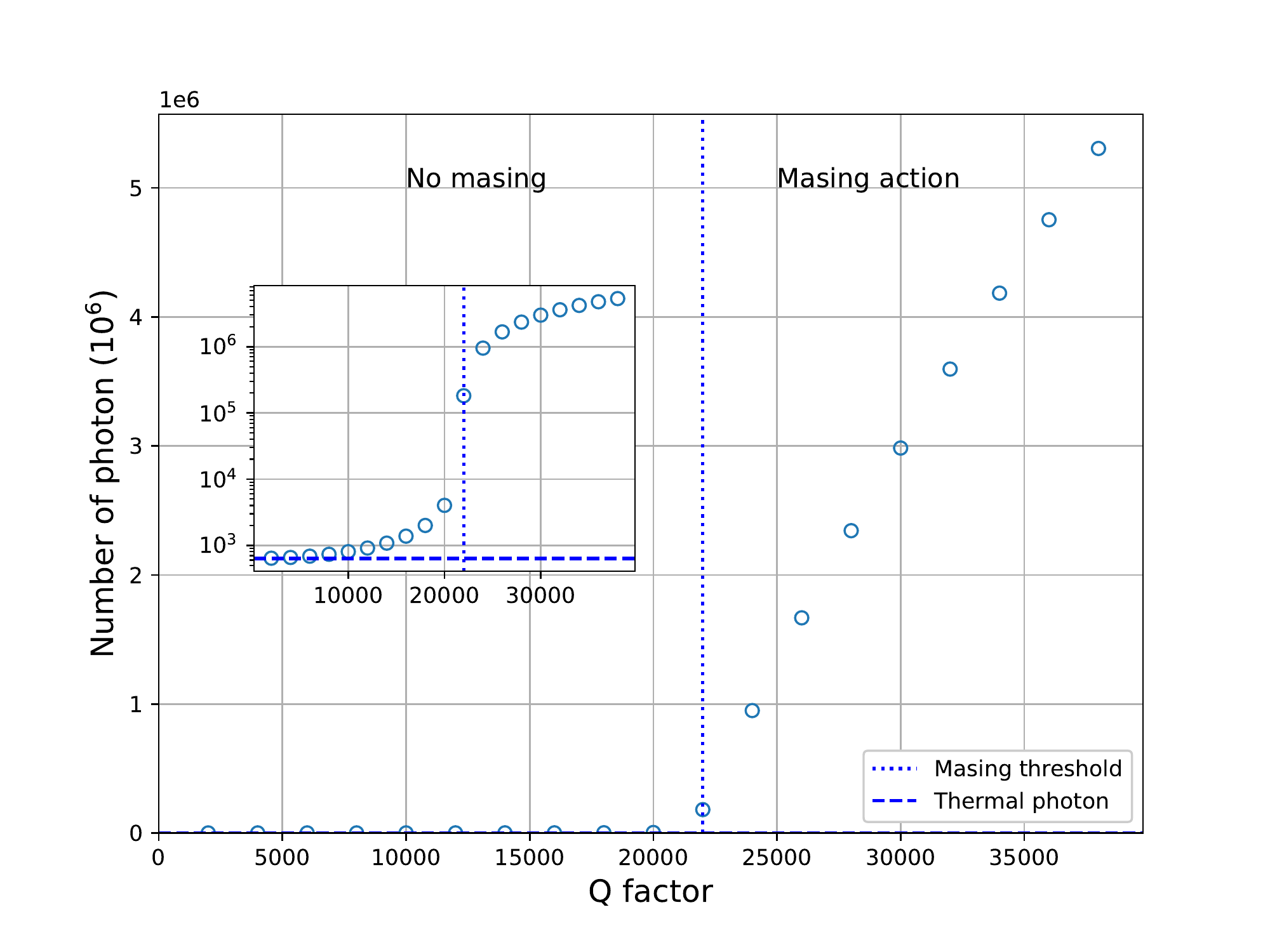}
    \caption{The steady-state number of photons as a function of the loaded cavity Q factor. The photon number shows threshold behaviour with respect to the $Q$ factor, with the onset of masing at around $Q=22000$ for the parameters listed in Table \ref{tab:Parameters}. The inset shows the onset on a logarithmic scale.}
    \label{fig:photonQ}
\end{figure}

\begin{figure}
    \centering
    \includegraphics[width=.9\columnwidth]{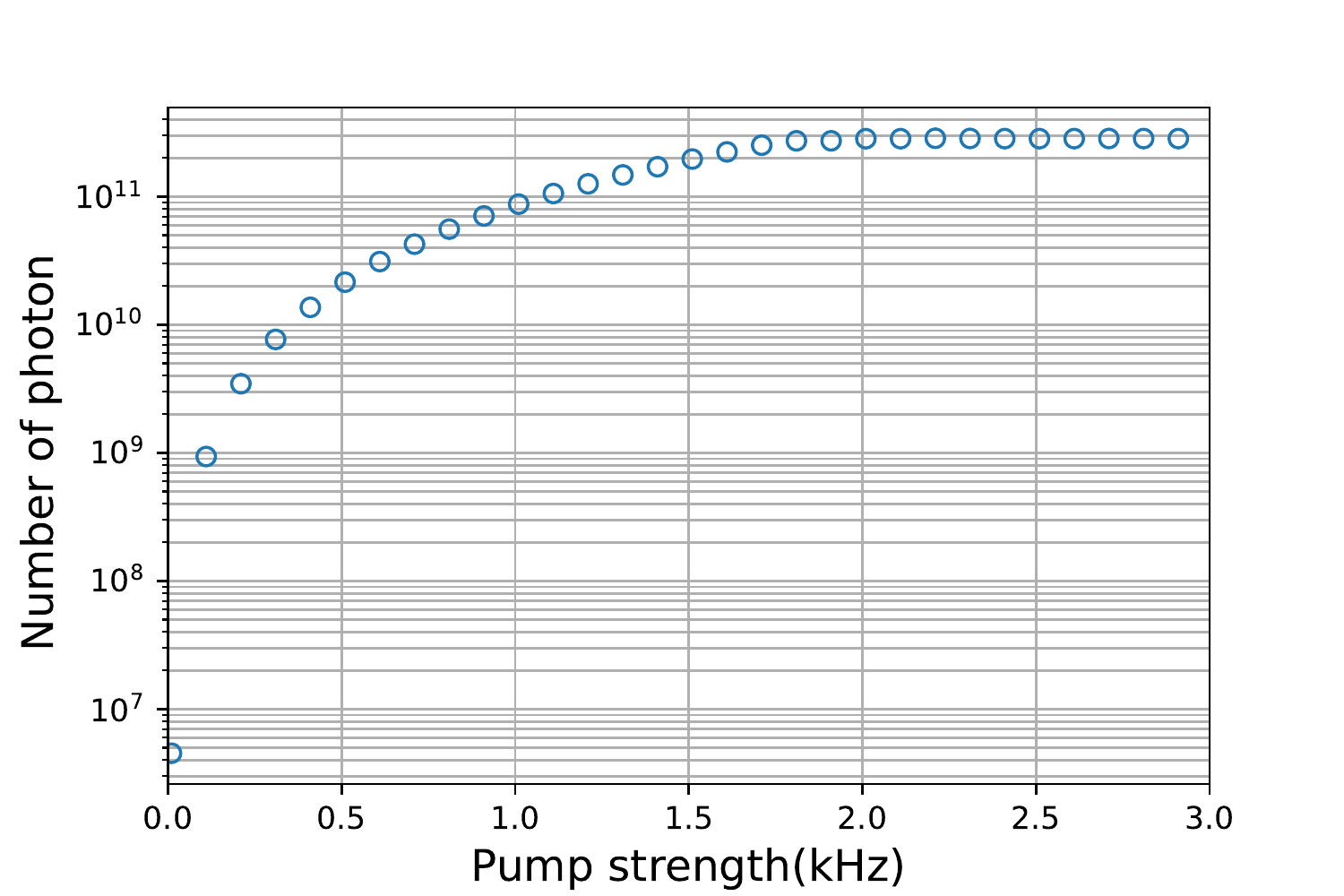}
    \caption{The steady-state number of photons as a function of the optical pump strength, defined by Eq (1), at low pump strengths. For the parameters listed in Table \ref{tab:Parameters}, the photon number saturates at a pump strength of around 1.5 kHz. }
    \label{fig:photonPump-Low}
\end{figure}
\begin{figure}
    \centering
    \includegraphics[width=.9\columnwidth]{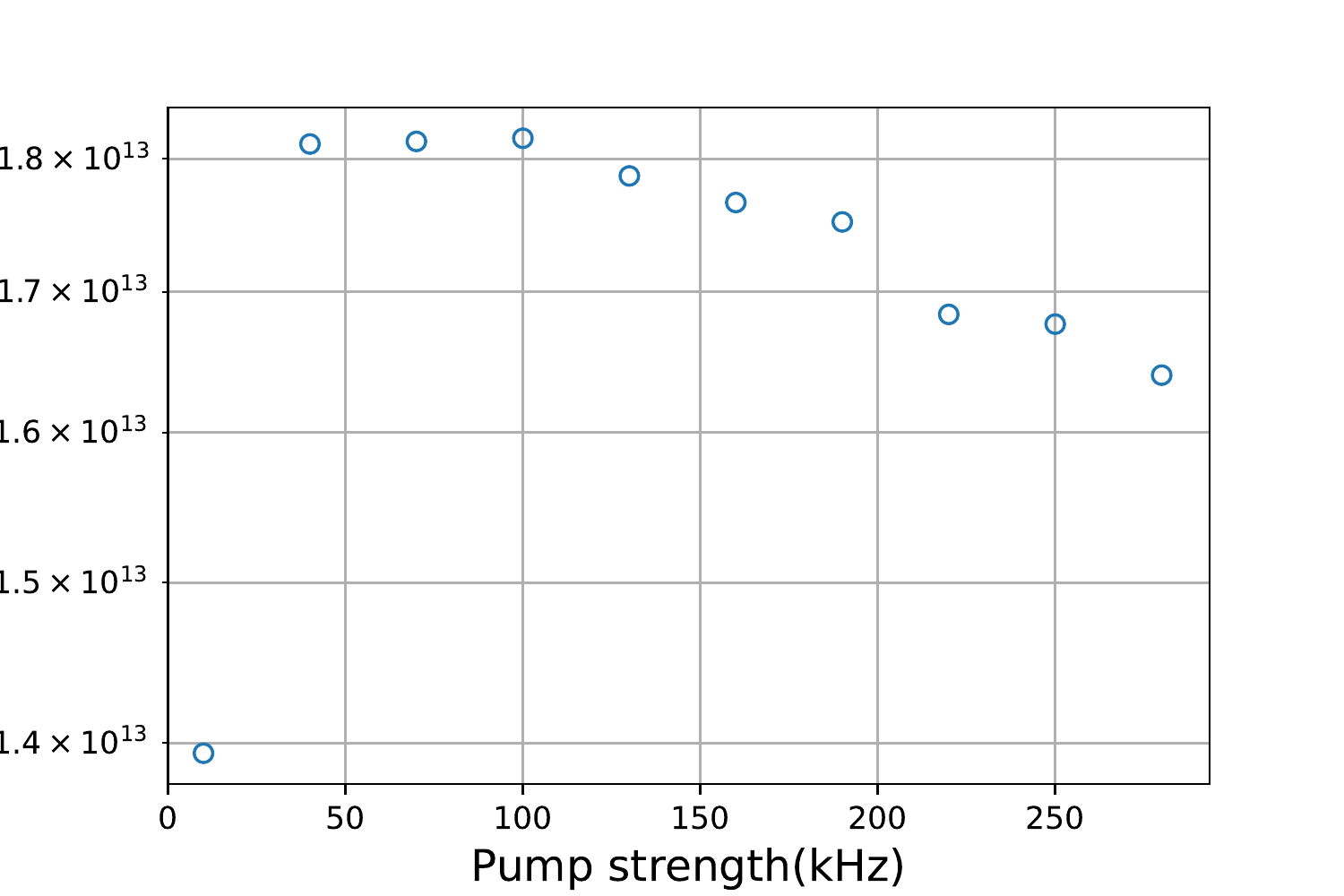}
    \caption{The steady-state number of photons as a function of the optical pump strength, defined by Eq. (1), at high pump strengths. For the parameters listed in Table \ref{tab:Parameters}, the number of microwave photons begins to decrease if the pump strength is increased beyond around 100 kHz because the NV$^{-}$ centres begin to accumulate in the $|7\rangle$ state due to its slow decay rate. This tends to reduce the population inversion and inhibit maser action. 
     }
    \label{fig:photonPump-High}
\end{figure}
\begin{figure}
    \centering
    \includegraphics[width=.9\columnwidth]{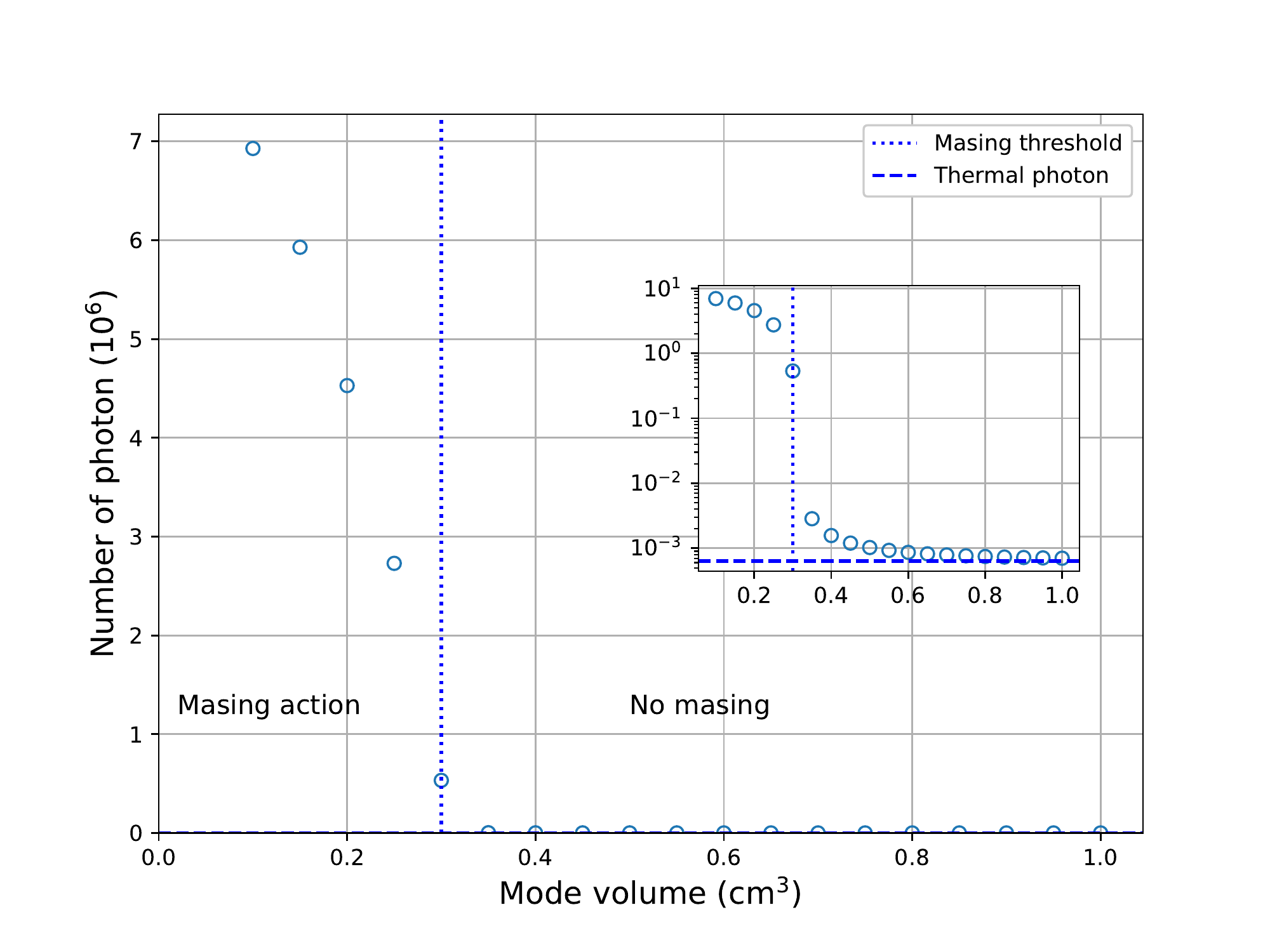}
    \caption{The steady-state number of microwave photons as a function of the cavity mode volume for the parameters listed in Table \ref{tab:Parameters}. The masing threshold occurs at a mode volume of about 0.3 cm$^3$. The inset shows the onset on a logarithmic scale.
    }
    \label{fig:photonModeV}
\end{figure}

\begin{figure}
    \centering
    \includegraphics[width=1.0\columnwidth]{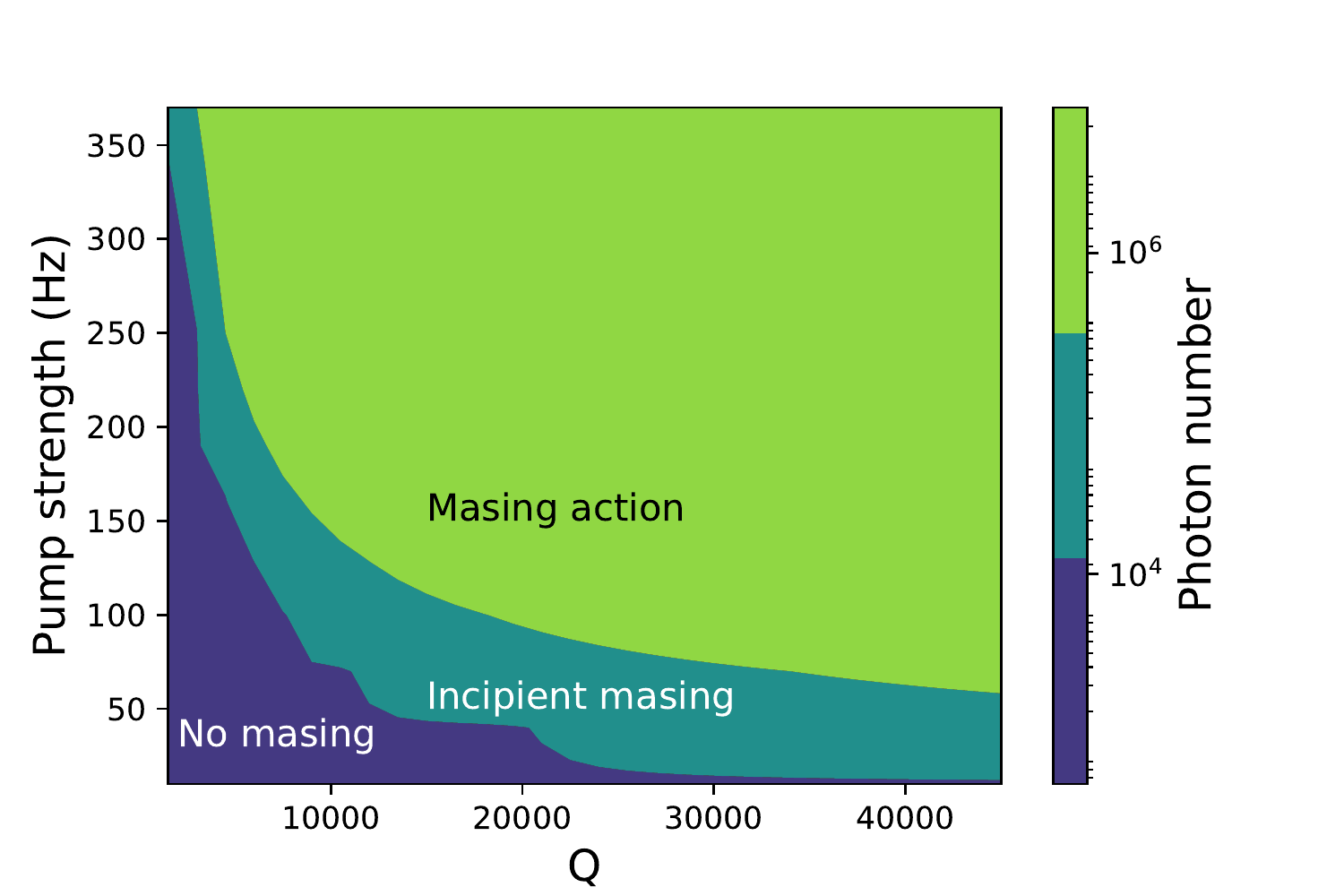}
    \caption{The steady-state microwave photon number as a function of the optical pump strength and the loaded cavity Q-factor. We identify a masing regime, a non-masing regime and an intermediate ``incipient masing'' regime in which stimulated emission dominates spontaneous emission, but the maser cannot achieve self-sustaining oscillations. At all pump strengths, the photon number increases monotonically with the Q-factor as expected.}
    \label{fig:qrmap}
\end{figure}
\begin{figure}
    \centering
    \includegraphics[width=1.0\columnwidth]{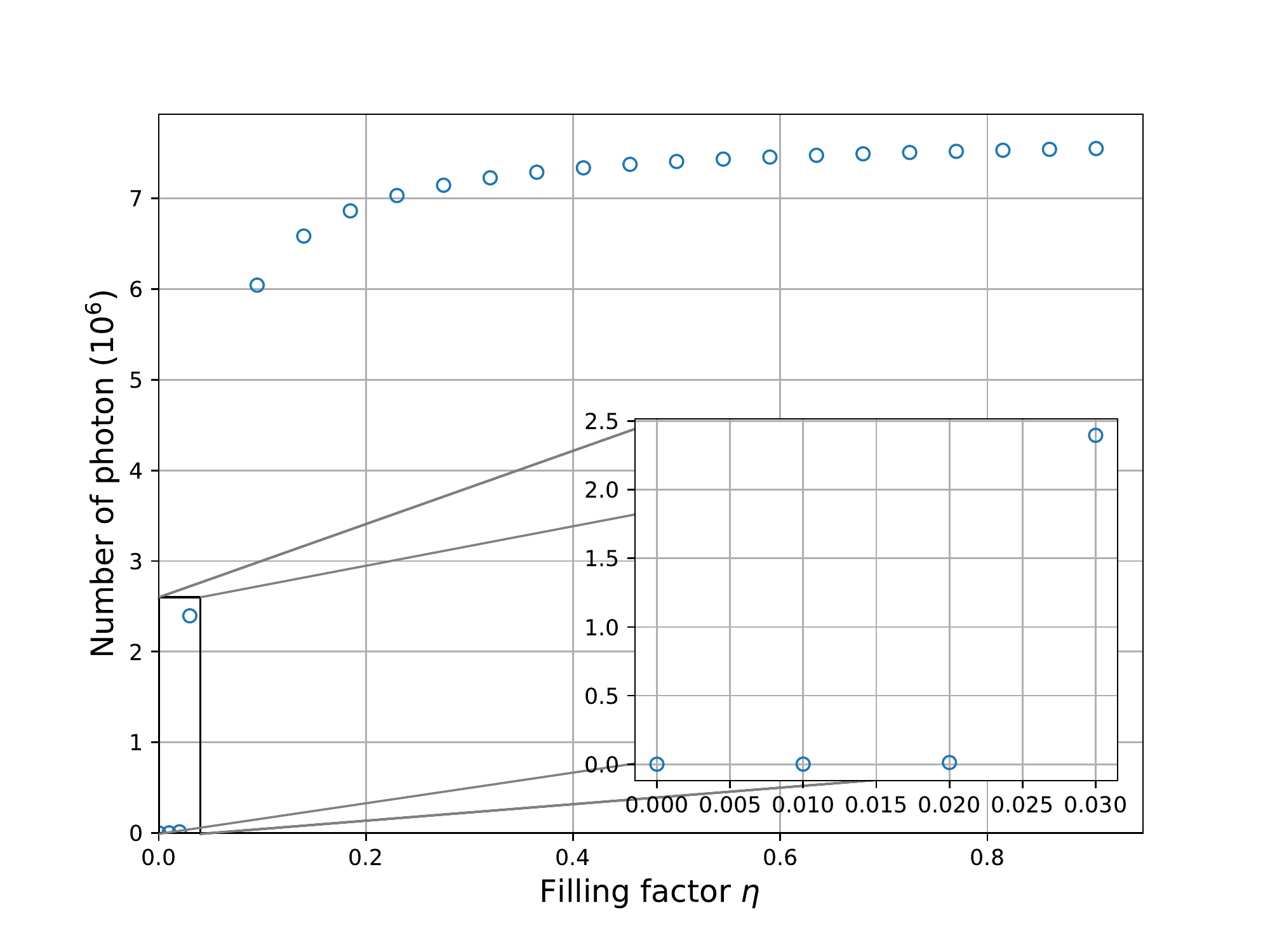}
    \caption{The number of microwave photons in the cavity as a function of the filling factor, defined by Eq. 4. The number of photons becomes saturated above $\eta=0.3$. The inset shows the onset of masing at a filling factor of 0.03.}
    \label{fig:fillingfactor}
\end{figure}

Looking at the intracavity photon number as a function of the loaded cavity Q-factor (Fig. 3), with other parameters fixed, we may identify a masing threshold as the point where the number of photons in the cavity is an order of magnitude greater than the thermal population. The maser output power in dBm can be estimated using $10\log_{10}((nh\omega_m^2/Q_m)/ 1\textrm{mW})$, where $\omega_m$ is the operating maser frequency, $h$ is the Planck constant, $n$ is the intracavity photon number and $Q_m$ is the Q factor for the external load, which can be estimated by $Q_m=1/(1/Q_l-1/Q_u)$, where $Q_l$ and $Q_u$ are the loaded and unloaded Q factors. Assuming that the loaded Q, which is the Q factor on the $x$-axis is 30000 and the unloaded Q is 50000, which is under coupled that the coupling constant is 0.67, the maser output is without any amplifiers is -108.5 dBm, calculated from the figure.
The plot in Fig. 3 starts with a low Q factor, the number of photons is close to the thermal photon number, which suggests that there is no masing. When the Q factor increases, the number of photons also increases. At a certain threshold Q factor, the number of photons increases significantly, where the slope of the plot is the greatest. At this point, the photon emission rate is higher than the cavity decay rate, thus, the number of photons in the cavity increase, and after this point, the maser can be maintained because a higher Q factor means that the cavity has a longer photon lifetime. 
The emission is predominantly stimulated in this case. 

In Figs. 4 and 5, the performance of the maser at different pump strengths is studied at a fixed Q factor of 30,000. It is clear that the maser operation becomes saturated when the pump strength is about 1 kHz, where the maser operation does not benefit from having a higher pump strength. If the pump power is increased further as in Fig. 5, the number of photons decreases after 100 kHz of excitation pump strength. At such high pump strength, the populations from the ground states are pumped to the excited state and accumulate in the singlet state $|7\rangle$, as its decay rate to the ground state $\kappa_{71}$ is slow compared to $\kappa_{63},\kappa_{52},\kappa_{41}$. Overall, the number of spins that participate in maser reduces at high pump strength and leads to a reduction in the number of photons in the cavity.

The threshold behaviour of the mode volume, which is inversely proportional to the Purcell factor and the cooperativity, is shown in Fig. 6. As expected, smaller mode volumes increase the number of masing photons in the cavity. For the parameters in Tables 1 and 2, the threshold is roughly at 0.3 cm$^3$, and for larger mode volumes masing cannot occur. 

Fig. 7 shows the number of masing photons as a function of the pump strength and the loaded Q factor. The dark blue region shows that the masing effect does not occur due to the high losses and low pump strength. The second region, in dark green, shows the ``incipient masing'' region, where stimulated emission starts to build up and the number of photons is higher than the number of thermal photons. In this region, an external MW signal fed into the maser would be amplified but the maser cannot be run as an oscillator. The maser is in operation in the light green region, where stimulated emission is maintained and the device can undergo self-sustaining maser oscillations.

The number of intracavity photons is shown as a function of the gain medium filling factor (defined by the ratio of the illuminated gain medium volume to the mode volume) in Fig. 8, again displaying a clear threshold for masing. The results show that the threshold is about 0.03, beyond which the number of photons maintained in the cavity significantly increases beyond the thermal population. The number of cavity photons increases monotonically as the filling factor approaches unity. 

It is assumed that the number of NV$^-$ centres is constant during masing. This is reasonable since the pump power required to achieve maser operation is low enough that the ionisation and recombination of NV centres due to laser excitation is weak \cite{cardoso2023impact,savvin2021nv}. Similarly, all the rates are assumed to be constant. In reality, the diamond will be heated by the laser, which will lead to a reduction in $T_1$ time \cite{zollitsch2023maser} as well as the $T_2$ time \cite{lin2021temperature}. Both coherence times vary with temperature as $T^5$, whereas the $T_2^*$ caused by the field inhomogeneity is only very weakly dependent on temperature \cite{lin2021temperature}. The zero-field splitting is also temperature dependent \cite{chen2011temperature}, but since this effect is also weak it is ignored in this model. All these temperature effects will tend to broaden the maser linewidth. A recent experiment has shown that the high power rate will significantly increase the temperature, especially the $T_1$ time, which makes maser action impossible \cite{zollitsch2023maser} by affecting the ability of the gain medium to maintain a population inversion. It would be interesting to develop the rate equation approach presented here to include such effects.

A further interesting extension to our rate equation model would be to explicitly model the effect of NV concentration on the T$_2$ time. To first order, changes in $NV^{-}$ concentration do not affect the cooperativity since the spin decay rate $\kappa_s$ is proportional to the $NV^{-}$ concentration\cite{wang2013spin} so that the factor of $N/\kappa_s$ in the cooperativity is constant. However, since the concentration of $NV^{-}$ in diamond samples depends not just on the concentration of nitrogen impurities but on the efficiency with which such impurities can be converted to nitrogen-vacancy centres and on the relative populations of the two possible charge states (NV$^{-}$ and NV$^{0}$), each of which can have an independent effect on the spin decay rate, the cooperativity is not independent of the NV$^{-}$ for real systems. The ratio of NV:NV$^-$ is typically\cite{breeze2018continuous} around 14:1, but a ratio of 6.95:1 has been reported for an optimised system\cite{edmonds2021characterisation}. Finally, we expect a higher concentration to lead to increased self-absorption \cite{ahn2007self}, which is not taken into account in our model but will again tend to increase the threshold pump power by increasing the effective cavity decay rate.

An important question is how field inhomogeneity affects diamond masers and to develop scalable room-temperature diamond masers it will be important to determine the minimum field inhomogeneity required. The homogeneous linewidth of the NV$^-$ resonance is about 1 MHz \cite{fortman2019understanding} and in these simulations, it is assumed that the spatial inhomogeneity of the applied magnetic field over the volume of the diamond is small enough that the resulting Zeeman shifts are within the homogeneous linewidths of the NV$^-$ centres. For an intrinsic linewidth of 1 MHz, this means the field inhomogeneity should be $< 0.03$ mT. In real applications, the smaller the field inhomogeneity the better the maser performance.

\section{Conclusion}

We have used extended second-order rate equations which incorporate the intracavity photon number and coherences between different NV energy states to simulate the spin dynamics of an ensemble of NV$^-$ centres in a diamond-based maser. The model predicts the behaviour of the maser as a function of material parameters of the gain medium and resonator in good agreement with recent experimental results \cite{zollitsch2023maser}.

Cooperativity and the related Purcell factor are the key quantities which, if increased, will reduce the threshold pump power for masing. On the one hand, this requires a high Q factor and a small mode volume, but intrinsic material parameters of the diamond gain media such as the concentration of NV$^-$ and the T$_2^*$ play a significant role, where a long $T_2^*$ and more excited spins, which are determined by the filling factor and the concentration of the sample, are favourable.  
High pump strength will eventually reduce the number of maser photons, as the populations end up in the singlet state and the population inversion is reduced due to a kinetic bottleneck leading to the accumulation of NV$^{-}$ centres in the $|7\rangle$ state. Increasing the transition rate from the $|7\rangle$ state through Purcell enhancement or by resonantly pumping transitions from the $|7\rangle$ state could thus be a pathway to higher maser power outputs.

Attention to these considerations will aid the design of diamond-based masers suitable for mass production. 

We further suggested improvements in the modelling of NV$^-$ ensembles for maser applications. If the NV$^-$ to NV$^0$ charge conversion and effects of temperature and NV$^-$ concentration on the spin lifetimes are considered, the trade-offs between maximising the coherence times and increasing output power and cooperativity by maximising the number of NV$^-$ centres and the pump power will become clearer. Such improvements will require further experimental input, particularly on the relationship between charge-state ratios and spin lifetimes.

The method presented in this paper, which computes the time evolution of the maser output explicitly, will be of particular use in the ``incipient masing'' regime where the stimulated emission is dominant over the spontaneous emission, but insufficient for self-sustaining maser oscillation and thus less amenable to solutions based on steady-state considerations. The full time-dependent response of diamond masers will also be useful for technological applications where masers are used to amplify arbitrary time-dependent microwave signals.

More broadly, the extended rate equation model presented in this paper will be of interest for modelling other quantum technologies based on ensembles of spin defects, such as diamond magnetometers \cite{barry2020sensitivity} and room-temperature masers based on alternative gain media such as silicon-vacancy defects in SiC \cite{kraus2014room,fischer2018highly, Gottscholl2022}.

\begin{acknowledgments}
\noindent This work was supported by the UK Engineering and Physical Sciences Research Council through the NAME Programme Grant (EP/V001914/1) and through grant EP/S000798/1. The authors further wish to acknowledge support from the Henry Royce Institute.
\\
\\
The authors have no conflicts to disclose.
\end{acknowledgments}

\bibliography{rates.bib}

\clearpage
\onecolumngrid
\appendix
\section{Matrix equations}

We present the rate equations solved numerically to generate the results of this work in matrix form. The overall equation of motion for the density matrix $\mathbf{\rho}$ is given by

\[\dot{\bm{\rho}}=\mathbf{M}\bm{\rho}\]
\begin{equation*}
\bm{\rho} = 
\begin{pmatrix}
\rho_{11} \\
\rho_{22} \\
\rho_{33} \\
\rho_{44} \\
\rho_{55} \\
\rho_{66} \\
\rho_{77} \\
\rho_{12} \\
\rho_{21} \\
\rho_{13} \\
\rho_{31} \\
\rho_{14} \\
\rho_{41} \\
\rho_{25} \\
\rho_{52} \\
\rho_{63} \\
\rho_{36} \\
\end{pmatrix}
\end{equation*}
\begin{equation*}
\mathbf{M} =
\begin{bmatrix}
\mathbf{M_{11}}&\mathbf{M_{12}}\\
\mathbf{M_{21}}&\mathbf{M_{22}}
\end{bmatrix}
\end{equation*}
where $\mathbf{M_{ij}}$ are sub-matrices of $\mathbf{M}$ used to clarify the equation by separating different contributions to the time evolution. $\mathbf{M_{11}}$ is the $7\times 7$ population sub-matrix, describing the spin dynamics of the 7-level model. $\mathbf{M_{12}}$ and $\mathbf{M_{21}}$ are the population-coupling sub-matrices, respectively $7\times 10$ and $10\times 7$, that describe the coupling of the laser field and the MW field with the 7-level model. Finally $\mathbf{M_{22}}$ is the coherence sub-matrix that describes the coherence of the coupled fields. This $17\times 17$ matrix contains 7 non-zero diagonal terms and 10 non-diagonal terms. The sub-matrices are given explicitly as
\begin{equation*}
\mathbf{M_{22}} =
\scriptsize{
\begin{bmatrix}
-\gamma_m+i \Delta_{m}&0&0&0&0&0&0&0&0&0\\
0&-\gamma_m+i \Delta_{m}&0&0&0&0&0&0&0&0\\
0&0&-\gamma_m+i \Delta_{m}&0&0&0&0&0&0&0\\
0&0&0&-\gamma_m+i \Delta_{m}&0&0&0&0&0&0\\
0&0&0&0&-\gamma_l+i \Delta_{l}&0&0&0&0&0\\
0&0&0&0&0&-\gamma_l+i \Delta_{l}&0&0&0&0\\
0&0&0&0&0&0&-\gamma_l+i \Delta_{l}&0&0&0\\
0&0&0&0&0&0&0&-\gamma_l+i \Delta_{l}&0&0\\
0&0&0&0&0&0&0&0&-\gamma_l+i \Delta_{l}&0\\
0&0&0&0&0&0&0&0&0&-\gamma_l+i \Delta_{l}

\end{bmatrix}}
\end{equation*}

\begin{equation*}
\mathbf{M_{11}} =
\begin{bmatrix}
-2\kappa_{21}&\kappa_{21}&\kappa_{21}&\kappa_{41}&0&0&\kappa_{71}\\
\kappa_{21} &-2\kappa_{21}&\kappa_{21}&0&\kappa_{52}&0&0\\
\kappa_{21}&\kappa_{21}&-2\kappa_{21}&0&0&\kappa_{63}&\kappa_{73}\\
0&0&0&-(\kappa_{41}+\kappa_{47})&0&0&0\\
0&0&0&0&-(\kappa_{52}+\kappa_{57})&0&0\\
0&0&0&0&0&-(\kappa_{63}+\kappa_{67})&0\\
0&0&0&\kappa_{47}&\kappa_{57}&\kappa_{67}&-(\kappa_{71}+\kappa_{72}+\kappa_{73})

\end{bmatrix}
\end{equation*}
\begin{equation*}
\mathbf{M_{12}} =
\begin{bmatrix}
-\dfrac{\Omega_m}{2}i&\dfrac{\Omega_m}{2}i&-\dfrac{\Omega_m}{2}i&\dfrac{\Omega_m}{2}i&-\dfrac{\Omega_l}{2}i&\dfrac{\Omega_l}{2}i&0&0&0&0 \\
\dfrac{\Omega_m}{2}i&-\dfrac{\Omega_m}{2}i&0&0&0&0&-\dfrac{\Omega_l}{2}i&\dfrac{\Omega_l}{2}i&0&0 \\
0&0&\dfrac{\Omega_m}{2}i&-\dfrac{\Omega_m}{2}i&0&0&0&0&-\dfrac{\Omega_l}{2}i&\dfrac{\Omega_l}{2}i \\
0&0&0&0&\dfrac{\Omega_l}{2}i&-\dfrac{\Omega_l}{2}i&0&0&0&0\\
0&0&0&0&0&0&\dfrac{\Omega_l}{2}i&-\dfrac{\Omega_l}{2}i&0&0\\
0&0&0&0&0&0&0&0&\dfrac{\Omega_l}{2}i&-\dfrac{\Omega_l}{2}i\\
0&0&0&0&0&0&0&0&0&0
\end{bmatrix}
\end{equation*}

\begin{equation*}
\mathbf{M_{21}} =
\begin{bmatrix}
-\dfrac{\Omega_m}{2}i&\dfrac{\Omega_m}{2}i&0&0&0&0&0\\
\dfrac{\Omega_m}{2}i&-\dfrac{\Omega_m}{2}i&0&0&0&0&0\\
-\dfrac{\Omega_m}{2}i&0&\dfrac{\Omega_m}{2}i&0&0&0&0\\
\dfrac{\Omega_m}{2}i&0&-\dfrac{\Omega_m}{2}i&0&0&0&0\\
-\dfrac{\Omega_m}{2}i&0&0&\dfrac{\Omega_m}{2}i&0&0&0\\
\dfrac{\Omega_m}{2}i&0&0&-\dfrac{\Omega_m}{2}i&0&0&0\\
-\dfrac{\Omega_l}{2}i&0&0&0&\dfrac{\Omega_l}{2}i&0&0\\
\dfrac{\Omega_l}{2}i&0&0&0&-\dfrac{\Omega_l}{2}i&0&0\\
0&-\dfrac{\Omega_l}{2}i&0&0&0&\dfrac{\Omega_l}{2}i&0\\
0&\dfrac{\Omega_l}{2}i&0&0&0&-\dfrac{\Omega_l}{2}i&0\\
0&0&-\dfrac{\Omega_l}{2}i&0&0&0&\dfrac{\Omega_l}{2}i\\
0&0&\dfrac{\Omega_l}{2}i&0&0&0&-\dfrac{\Omega_l}{2}i

\end{bmatrix}
\end{equation*}
where $\gamma_m$ is the dephasing rate of the MW-coupled spin states, $\gamma_l$ is the decoherence rate of the laser-coupled states and $\Delta_l$ is the detuning of the laser from the transition frequencies for $1\to 4$, $2\to5$ and $3\to 6$. For the results presented in this work, the laser was modelled as resonantly exciting the ground states to the excited states so that the detuning, $\Delta_l$, is zero. $\Delta_m$ is the detuning of the microwave from the maser operating frequency. 

In the extended rate equations, an additional second-order term, representing the number of photons, is added to the first-order rate equations.
\begin{equation*}
\dfrac{da}{dt}=\kappa_{sp}\rho_{11}+\kappa_{st}(\rho_{11}-\rho_{22})a-\kappa_c(n_{th}-a)
\end{equation*}
where $\kappa_{sp}$ is the spontaneous emission rate and $\kappa_{st}$ is the stimulated emission rate. 
\clearpage
\end{document}